\documentclass[%
mathleft,%
]{an}
\usepackage{graphicx}
\usepackage{times}
\usepackage{natbib}
\begin{document}

\Pagespan{1}{}
\Yearpublication{2011}%
\Yearsubmission{2011}%
\Month{1}%
\Volume{999}%
\Issue{92}%

\title{Statistical properties of twin kHz QPO in neutron star LMXBs}

\author{D.\,H. Wang\inst{1,2}
\and L. Chen\inst{1}
\and C.\,M. Zhang\inst{2}\fnmsep\thanks{Corresponding author: zhangcm@bao.ac.cn}
\and Y.\,J. Lei\inst{2}
\and J.\,L. Qu\inst{3}
}
\titlerunning{Statistical properties of twin kHz QPO in neutron star LMXBs}
\authorrunning{D.\,H. Wang et al.}
\institute{
Astronomy Department, Beijing Normal University, Beijing 100875
\and
National Astronomical Observatories, University of Chinese Academy of Sciences, Beijing 100012, China
\and
Institute of High Energy Physics, University of Chinese Academy of Sciences, Beijing 100049, China
}

\received{XXXX}
\accepted{XXXX}
\publonline{XXXX}


\abstract{%
We collect the data of twin kilohertz quasi-periodic oscillations (kHz QPOs) published before 2012 from 26 neutron star (NS)
low-mass X-ray binary (LMXB) sources, then we analyze the centroid
frequency ($\nu$) distribution of twin kHz QPOs (lower frequency $\nu_1$ and
upper frequency $\nu_2$) both for Atoll and Z sources. For the data without shift-and-add, we find that Atoll and Z sources show the different distributions of $\nu_1$, $\nu_2$ and $\nu_2/\nu_1$, but the same distribution of $\Delta\nu$ (difference of twin kHz QPOs), which indicates that twin kHz QPOs may share the common
properties of LXMBs and have the same physical origins. The distribution of $\Delta\nu$ is quite different from constant value, so is $\nu_2/\nu_1$ from constant ratio.
The weighted mean values and maxima of $\nu_1$ and $\nu_2$ in Atoll sources are
slightly higher than those in Z sources.
We also find that shift-and-add technique can reconstruct the distribution of $\nu_1$ and $\Delta\nu$. The K-S test results of $\nu_1$ and $\Delta\nu$ between Atoll and Z sources from data with shift-and-add are quite different from those without it, and we think that this may be caused by the selection biases of the sample.
We also study the properties of the quality factor ($Q$) and the root-mean-squared
(rms) amplitude of 4U 0614+09 with the data from the two observational methods, but the errors are too big to make a robust conclusion. The NS spin frequency ($\nu_s$) distribution of 28 NS-LMXBs show a bigger mean value ($\sim$ 408\,Hz) than that
($\sim$ 281\,Hz) of the radio binary millisecond pulsars (MSPs), which may be due to the lack of the spin detections
from Z sources (systematically lower than 281\,Hz).
Furthermore, on the relations between the kHz QPOs and NS spin
frequency $\nu_s$, we find the approximate correlations of the mean values of $\Delta\nu$ with NS spin and its half, respectively.
}

\maketitle

\section{Introduction}
\begin{center}
\begin{table*}
\begin{minipage}[]{170mm}
\caption[]{Twin kHz QPOs (without shift-and-add)}
\begin{tabular}{@{}lccccl@{}}
\noalign{\smallskip}\hline\noalign{\smallskip}
Source (22) & $\nu_1$& $\langle\nu_1\rangle$ & $\nu_2$ & $\langle\nu_2\rangle$ & Ref \\
 & (Hz) & (Hz) & (Hz) & (Hz) & \\
\hline\noalign{\smallskip}
Millisecond pulsars (2) \\
SAX J1808.4-3658 & $499\pm4\sim503.6\pm5.3$ & $501\pm2$ & $685.1\pm5.1\sim694\pm4$ & $691\pm4$ & 1 \\
XTE J1807.4-294 & $106\pm30\sim370\pm18$ & $237\pm16$ & $337\pm10\sim587.0\pm4.1$ & $437\pm15$ & 2 \\
\noalign{\smallskip}\hline\noalign{\smallskip}
Atoll Sources (12) \\
4U 0614+09 & $153.4\pm5.6\sim823.7\pm10.0$ & $651\pm18$ & $449.4\pm19.5\sim1161.8\pm4.6$ & $944\pm21$ & 3 \\
4U 1608-52 & $531\pm17\sim784.35\pm0.91$ & $719\pm48$ & $830.3\pm5.8\sim1061.9\pm6.3$ & $918\pm39$ & 4 \\
4U 1636-53 & $565.4\pm5.1\sim921.9\pm7.7$ & $839\pm22$ & $860.4\pm1.7\sim1194\pm19$ & $944\pm33$ & 5 \\
4U 1728-34 & $305\pm8\sim879.2\pm3.0$ & $740\pm20$ & $582\pm10\sim1161\pm16$ & $892\pm14$ & 6 \\
4U 1735-44 & $640.5\pm2.5\sim900\pm14$ & $727\pm6$ & $981.7\pm6.7\sim1149\pm4$ & $1097\pm43$ & 7 \\
4U 1820-30 & $764\pm6\sim796\pm4$ & $789\pm5$ & $1055\pm10\sim1072\pm11$ & $1067\pm4$ & 8 \\
4U 1915-05 & $223.6\pm4.7\sim706.9\pm22.4$ & $440.6\pm85.8$ & $513.6\pm0.2\sim1055.3\pm17.9$ & $513.8\pm5.2$ & 9 \\
IGR J17511-3057$^\dag$ & $72.5\pm4.9\sim140.3\pm3.4$ & $124.9\pm13.4$ & $179.9\pm14.9\sim272.2\pm13.9$ & $247.7\pm17.3$ & 10 \\
KS 1731-260 & $898.3\pm3.3\sim903.3\pm2.7$ & $900.7\pm1.1$ & $1158.6\pm9.0\sim1176.2\pm2.9$ & $1173.5\pm3.0$ & 11 \\
SAX J1750.8-2900 & $936\pm1$ & $936\pm1$ & $1253\pm9$ & $1253\pm9$ & 12 \\
XTE J1701-407 & $745\pm9$ & $745\pm9$ & $1150\pm7$ & $1150\pm7$ & 13 \\
XTE J2123-058 & $847.1\pm5.5\sim871\pm2$ & $859\pm6$ & $1102\pm13\sim1141\pm5$ & $1131\pm8$ & 14 \\
\noalign{\smallskip}\hline\noalign{\smallskip}
Z Sources (8) \\
Cir X-1$^\ddag$ & $56.1\pm1.3\sim226\pm18$ & $82\pm9$ & $229\pm18\sim505\pm51$ & $358\pm26$ & 15 \\
Cyg X-2 (2142+380) & $516\pm27$ & $516\pm27$ & $862\pm11$ & $862\pm11$ & 16 \\
GX 5-1 (1758-250) & $156\pm23\sim662\pm21$ & $280\pm23$ & $478\pm15\sim888\pm24$ & $671\pm18$ & 17 \\
GX 17+2 (1813-140) & $475\pm7\sim830\pm19$ & $641\pm24$ & $759\pm5\sim1079\pm1$ & $929\pm24$ & 18 \\
GX 340+0 (1642-455) & $197\pm70\sim565\pm12$ & $357\pm22$ & $535\pm85\sim840\pm21$ & $681\pm21$ & 19 \\
GX 349+2 (1702-363) & $715\pm12$ & $715\pm12$ & $985\pm7$ & $985\pm7$ & 20 \\
Sco X-1 (1617-155) & $565\pm4\sim853\pm5$ & $645\pm9$ & $872\pm2\sim1080\pm3$ & $943\pm9$ & 21 \\
XTE J1701-462$^\ddag$ & $502.4\pm23.1\sim650.5\pm7.1$ & $629\pm5$ & $760.8\pm6.4\sim944.9\pm3.7$ & $910\pm14$ & 22 \\
\noalign{\smallskip}\hline\noalign{\smallskip}
\end{tabular}
\end{minipage}
\begin{tabular}{@{}l@{}}
\begin{minipage}[]{165mm}
$\dag$: The identification of the QPOs is uncertain. \\
$\ddag$: The source shows the properties of both Atoll and Z sources. \\
1. \citealt{van Straaten05},  \citealt{Wijnands03};
2. \citealt{Linares05}, \citealt{Zhang06b};
3. \citealt{van Straaten00}, \citealt{van Straaten02};
4. \citealt{van Straaten03};
5. \citealt{Altamirano08b}, \citealt{Wijnands97a}, \citealt{Bhattacharyya10};
6. \citealt{Di Salvo01}, \citealt{van Straaten02}, \citealt{Strohmayer96}, \citealt{Migliari03};
7. \citealt{Wijnands98c}, \citealt{Ford98b};
8. \citealt{Smale97};
9. \citealt{Boirin00};
10. \citealt{Kalamkar11};
11. \citealt{Wijnands97};
12. \citealt{Kaaret02};
13. \citealt{Strohmayer08b};
14. \citealt{Tomsick99}, \citealt{Homan99};
15. \citealt{Boutloukos06};
16. \citealt{Wijnands98a};
17. \citealt{Wijnands98b}, \citealt{Jonker02b};
18. \citealt{Homan02}, \citealt{Wijnands97b};
19. \citealt{Jonker02b}, \citealt{Wijnands98a}, \citealt{Jonker98};
20. \citealt{O'Neill02};
21. \citealt{van der Klis97}, \citealt{van der Klis96};
22. \citealt{Homan07}, \citealt{Homan10}, \citealt{Sanna10}.
\end{minipage}
\end{tabular}
\end{table*}
\end{center}
\begin{table*}
\begin{minipage}[]{170mm}
\caption[]{Twin kHz QPOs (with shift-and-add)}
\setlength{\tabcolsep}{2pt}
\begin{tabular}{@{}lccccl@{}}
\noalign{\smallskip}\hline\noalign{\smallskip}
Source (9) & $\nu_1$& $\langle\nu_1\rangle$ & $\Delta\nu$ & $\langle\Delta\nu\rangle$ & Ref \\
 & (Hz) & (Hz) & (Hz) & (Hz) & \\
\noalign{\smallskip}\hline\noalign{\smallskip}
Atoll Sources (8) \\
4U 0614+09 & $560.1\pm2.1\sim842.9\pm19.8$ & $644.1\pm9.7$ & $300.0\pm11.8\sim347.9\pm14.7$ & $320.5\pm2.3$ & 1 \\
4U 1608-52 & $473\sim867$ & $754\pm15$ & $225\pm12\sim326\pm3$ & $305\pm3$ & 2 \\
4U 1636-53 & $528.56\pm16.54\sim978.92\pm9.37$ & $841\pm7$ & $229.86\pm12.38\sim341.02\pm15.29$ & $269\pm6$ & 3 \\
4U 1702-43 & 722 & 722 & $333\pm5$ & $333\pm5$ & 4 \\
4U 1705-44 & $776.1\pm3.9$ & $776.1\pm3.9$ & $298.1\pm11.1$ & $298.1\pm11.1$ & 5 \\
4U 1728-34 & $576\sim894$ & $746$ & $279\pm12\sim356\pm8$ & $346\pm3$ & 6 \\
Aql X-1 (1908+005) & $795.45\pm0.04\sim803.09\pm0.05$ & $798.43\pm3.73$ & $278.1\pm18.3\sim280.1\pm13.4$ & $279.4\pm1.0$ & 7 \\
IGR J17191-2821 & $681\pm5\sim879\pm1$ & $793\pm42$ & $315\pm50\sim362\pm11$ & $349\pm7$ & 8 \\
\noalign{\smallskip}\hline\noalign{\smallskip}
Z Sources (1) \\
Sco X-1 (1617-155) & $531.6\pm16.6\sim902.2\pm0.1$ & $837\pm6$ & $240.6\pm4.4\sim337.5\pm14.7$ & $293\pm2$ & 9 \\
\noalign{\smallskip}\hline\noalign{\smallskip}
\end{tabular}
\end{minipage}
\begin{tabular}{@{}l@{}}
\begin{minipage}{165mm}
1. \citealt{Boutelier09};
2. \citealt{Barret05a}, \citealt{Jonker00a}, \citealt{Mendez98b};
3. \citealt{Di Salvo03}, \citealt{Jonker00a}, \citealt{Jonker02a}, \citealt{Lin11};
4. \citealt{Markwardt99};
5. \citealt{Ford98a};
6. \citealt{Jonker00a}, \citealt{Mendez99};
7. \citealt{Barret08};
8. \citealt{Altamirano10a};
9. \citealt{Lin11}, \citealt{Mendez00}.
\end{minipage}
\end{tabular}
\end{table*}
\begin{table*}
\begin{minipage}[]{170mm}
\caption[]{Twin kHz QPOs (with and without shift-and-add)}
\setlength{\tabcolsep}{2pt}
\begin{tabular}{@{}lccccl@{}}
\noalign{\smallskip}\hline\noalign{\smallskip}
Source (4)$\dag$ & $\nu_2$ & $\langle\nu_2\rangle$ & $\Delta\nu$ & $\langle\Delta\nu\rangle$  \\
 & (Hz) & (Hz)) \\
\noalign{\smallskip}\hline\noalign{\smallskip}
Without shift-and-add \\
4U 0614+09 & $449.4\pm19.5\sim1161.8\pm4.6$ & $944\pm21$ & $238.3\pm6.7\sim382.3\pm7.4$ & $316\pm5$ \\
4U 1608-52 & $830.3\pm5.8\sim1061.9\pm6.3$ & $918\pm39$ & $277.6\pm6.4\sim305\pm7$ & $293\pm5$ \\
4U 1636-53 & $860.4\pm1.7\sim1194\pm19$ & $944\pm33$ & $249.5\pm13.0\sim319$ & $296\pm5$ \\
4U 1728-34 & $582\pm10\sim1161\pm16$ & $892\pm14$ & $231\pm21\sim362.8\pm6.0$ & $321\pm7$ \\
Sco X-1 & $872\pm2\sim1080\pm3$ & $943\pm9$ & $223.1\pm5.3\sim312.1\pm3.1$ & $295\pm2$ \\
\noalign{\smallskip}\hline\noalign{\smallskip}
With shift-and-add \\
4U 0614+09 & $889.4\pm10.5\sim1144.4\pm6.8$ & $1002\pm20$ & $300.0\pm11.8\sim347.9\pm14.7$ & $320.5\pm2.3$ \\
4U 1608-52 & $799\pm3\sim1103.9\pm17.9$ & $952\pm14$ & $225\pm12\sim326\pm3$ & $305\pm3$ \\
4U 1636-53 & $822.9\pm2.5\sim1227.7\pm2.7$ & $1078\pm27$ & $229.86\pm12.38\sim341.02\pm15.29$ & $269\pm6$ \\
4U 1728-34 & $925.3\sim1183.2$ & $1085\pm25$ & $279\pm12\sim356\pm8$ & $346\pm3$ \\
Sco X-1 & $842.1\pm2.8\sim1142.8\pm4.4$ & $947\pm12$ & $240.6\pm4.4\sim337.5\pm14.7$ & $293\pm2$ \\
\noalign{\smallskip}\hline\noalign{\smallskip}
\end{tabular}
\end{minipage}
\begin{tabular}{@{}l@{}}
\begin{minipage}{165mm}
$\dag$ The reference are the same as Table 1 and 2.
\end{minipage}
\end{tabular}
\end{table*}
Kilohertz quasi-periodic oscillations (kHz QPOs) in neutron star
(NS) low mass X-ray binaries (LMXBs) often occur in pairs and have
been detected in both Atoll and Z sources \cite{van der Klis00,van der Klis06}. These frequencies cover the range from
several hundred Hertz to more than one kHz, where
\citet{Bhattacharyya10} has reported a high kHz QPO frequency of
an approximate 1860\,Hz, which may be the overtone of upper or
lower kHz QPOs. Twin kHz QPOs have the rather coherent statisticale
behaviors, and correlate with low frequency QPOs \citep{Belloni02,Psaltis99a,Psaltis99b}, so these phenomena are used as the powerful tool to
probe the physical process around NS-LMXBs \citep{van der
Klis06,van der Klis08}.

The kHz QPOs are the peaks superposed on the noise in the power
density spectra (PDS), and their profiles can be described by the
Lorentzian function \citep{van der Klis06}:
\begin{equation}
P_{\nu}\propto\lambda/[(\nu-\nu_0)^2+(\lambda/2)^2],
\end{equation}
where $\nu_0$ is the centroid frequency, $\lambda$ is the full width at
half-maximum (FWHM). The ratio of $\nu_0$ to $\lambda$ is defined as the quality factor:
\begin{equation}
Q\equiv\nu_0/\lambda,
\end{equation}
where the signals with $Q>2$ are considered as the QPO signals
while those with $Q<2$ are considered as the peaked noises. The
strength of a signal is described by its fractional
root-mean-squared (rms) amplitude, which is proportional to the
integrated power of its contribution to the power spectrum and
often expressed in percent. So the twin kHz QPOs in NS-LMXBs have
three characteristic parameters: the centroid frequency (lower
$\nu_1$, upper $\nu_2$), their corresponding quality factors
($Q_1$, $Q_2$) and fractional root-mean-squared amplitudes
($rms_1$, $rms_2$).

Some authors raised their models to explain the correlation between $\nu_1$ and $\nu_2$: Stella et al.
\citep{Stella99a,Stella99b} presented the relativistic precession model,
which can be applied to estimate the NS mass \citep{Torok10}. But, the fits of the model to the observational data of 4U 1636-53 and Sco X-1, are not satisfactory because of the large chi-squares \citep{Lin11,Torok08,Torok12}.
\citet{Zhang04} put forward the Alfv\'en wave oscillation model,
which can constrain NS $M-R$ relation \citep{Zhang07,Zhang09}.
\citet{Stuchlik08,Stuchlik11} studied the behavior of the effective gravitational potential around the specific resonant radii, and tried to explain the high-frequency QPOs as the forced resonant oscillations excited by gravitational perturbations. They proposed the resonant switch model \citep{Stuchlik12,Stuchlik13}, assuming a resonant point switch of one pair of the oscillation modes to some others due to non-linear resonant phenomena, which fits well with the data of 4U 1636-53.
Some researchers  proposed the sonic-point and spin-resonance models \citep{Miller98,Lamb01,Lamb03}, which explain that the frequency separation of the twin kHz QPOs is close to the spin frequency of NS or its half.
\citet{Abramowicz03a,Abramowicz03b,Kluzniak01}
introduced the resonance model with the upper frequency corresponding to the vertical epicyclic frequency and the lower frequency corresponding to the radial epicyclic frequency, which accords with observational data quite well in the black-hole binary systems \citep{Torok05,Torok11}.
But the last two
models show the inconsistent with the subsequent twin kHz QPO observations \citep{Mendez99,Jonker02a,Muno04,Strohmayer06,Wijnands03,van der Klis06,Zhang06a,Belloni07,Urbanec10}.

Besides the correlation between the upper and lower kHz QPOs,
many works have been dedicated to investigate the relation between
the quality factor and the centroid frequency of twin kHz QPOs. In
a series of work by
\citet{Barret05a,Barret05b,Barret05c,Barret06,Barret07,Barret08,Barret11},
the authors study the $Q-\nu$ relation among source 4U 1636-536,
4U 1608-522, 4U 1735-44, 4U 1728-34, 4U 1820-303, 4U 0614+09 and
XTE J1701.462. They find that all sources except 4U 0614+09 show
an abrupt drop in $Q_1~vs.~\nu_1$ plot. At the same time, $Q_2$ in
source 4U 1636-536, 4U 1608-522, 4U 1735-44, 4U 1820-303 and
4U 0614+09 increase all the way with  $\nu_2$. There are also some
similar investigations
\citep{Boutelier09,Boutelier10,Mendez06,Torok09}. \citet{Wang12}
study the quality factors of twin kHz QPOs in a statistical way,
and they find that the range of $Q$ in Atoll sources is wider than
those in Z sources for both upper and lower kHz QPOs.

In this paper,
we analyze the the centroid frequency distributions of twin kHz QPOs in Atoll and Z sources, then investigate the data from the
different observational methods (shift-and-add or not, see \citealt{Mendez98a}). As an extension of studying the influence of the two observational methods on kHz QPOs, we take 4U 0614+09 as an example to analyze the quality factor and rms amplitude.
The spin frequency distribution of NS-LMXBs are analyzed, which
is compared to that of the binary radio millisecond pulsars
(MSPs).
The whole pictures of kHz QPOs of both Atoll and Z sources are
discussed and pointed out.

\section{Parameter analysis for twin kHz QPOs}

\begin{table}
\begin{minipage}{80mm}
\caption{K-S test result ($\alpha=0.05$)}
\begin{tabular}{@{}ll@{}}
\noalign{\smallskip}\hline\noalign{\smallskip}
Parameter & p-value\\
\noalign{\smallskip}\hline\noalign{\smallskip}
Atoll vs. Z (without shift-and-add) & \\
$\nu_1$ & $6.1\times10^{-8}$ \\
$\nu_2$ & $5.8\times10^{-11}$ \\
$\Delta\nu$ & $7.4\times10^{-2}$ \\
$\nu_2/\nu_1$ & $4.3\times10^{-5}$ \\
\noalign{\smallskip}\hline\noalign{\smallskip}
Atoll vs. Z (with shift-and-add) & \\
$\nu_1$ & $4.1\times10^{-1}$ \\
$\Delta\nu$ & $1.0\times10^{-6}$ \\
\noalign{\smallskip}\hline\noalign{\smallskip}
Atoll vs. Z (with and without shift-and-add) & \\
$\nu_1$ & $1.4\times10^{-8}$ \\
$\nu_2$ & $1.4\times10^{-12}$ \\
$\Delta\nu$ & $7.4\times10^{-6}$ \\
$\nu_2/\nu_1$ & $8.4\times10^{-5}$ \\
\noalign{\smallskip}\hline\noalign{\smallskip}
With vs. without shift-and-add (4U 0614+09$^\dag$) & \\
$\nu_1$ & $1.2\times10^{-1}$ \\
$\Delta\nu$ & $6.2\times10^{-1}$ \\
\noalign{\smallskip}\hline\noalign{\smallskip}
With vs. without shift-and-add (Sco X-1$^\ddag$) & \\
$\nu_1$ & $3.6\times10^{-1}$ \\
$\Delta\nu$ & $3.6\times10^{-1}$ \\
\noalign{\smallskip}\hline\noalign{\smallskip}
With vs. without shift-and-add (Atoll) & \\
$\nu_1$ & $1.8\times10^{-1}$ \\
$\Delta\nu$ & $5.0\times10^{-1}$ \\
\noalign{\smallskip}\hline\noalign{\smallskip}
With vs. without shift-and-add (Z) & \\
$\nu_1$ & $1.9\times10^{-8}$ \\
$\Delta\nu$ & $4.1\times10^{-2}$ \\
\noalign{\smallskip}\hline\noalign{\smallskip}
$\Delta\nu$ (Atoll, without shift-and-add) vs. 300\,Hz & $1.4\times10^{-14}$ \\
$\Delta\nu$ (Z, without shift-and-add) vs. 300\,Hz & $7.9\times10^{-18}$ \\
\noalign{\smallskip}\hline\noalign{\smallskip}
$\nu_2/\nu_1$ (Atoll, without shift-and-add) vs. 1.5 & $1.2\times10^{-15}$ \\
$\nu_2/\nu_1$ (Z, without shift-and-add) vs. 1.5 & $9.2\times10^{-20}$ \\
\noalign{\smallskip}\hline\noalign{\smallskip}
NS spin in LMXBs vs. in Binary radio MSPs & \\
$\nu_s$ & $1.2\times10^{-3}$ \\
\noalign{\smallskip}\hline\noalign{\smallskip}
\end{tabular}
\end{minipage}
\\
\begin{tabular}{@{}l@{}}
\begin{minipage}{80mm}
$^\dag$\citealt{van Straaten00} and \citealt{Boutelier09}; \\
$^\ddag$\citealt{van der Klis97} and \citealt{Mendez00}.
\end{minipage}
\end{tabular}
\end{table}

We collect the data of twin kHz QPO from 26 NS-LMXB sources\footnote{The detail information of the sources can be seen
in \citet{Liu07}.}
published before 2012, which includes 2 millisecond pulsars, 16 Atoll
sources, 8 Z sources. In some cases, no tables of kHz QPO are provided, and we obtain the data from authors or figures. The identification of the QPOs in IGR J17511-3057 is uncertain \citep{Kalamkar11}, so we do not use this data when analyzing the distribution of Atoll sources.
\begin{figure*}
\includegraphics[width=8cm]{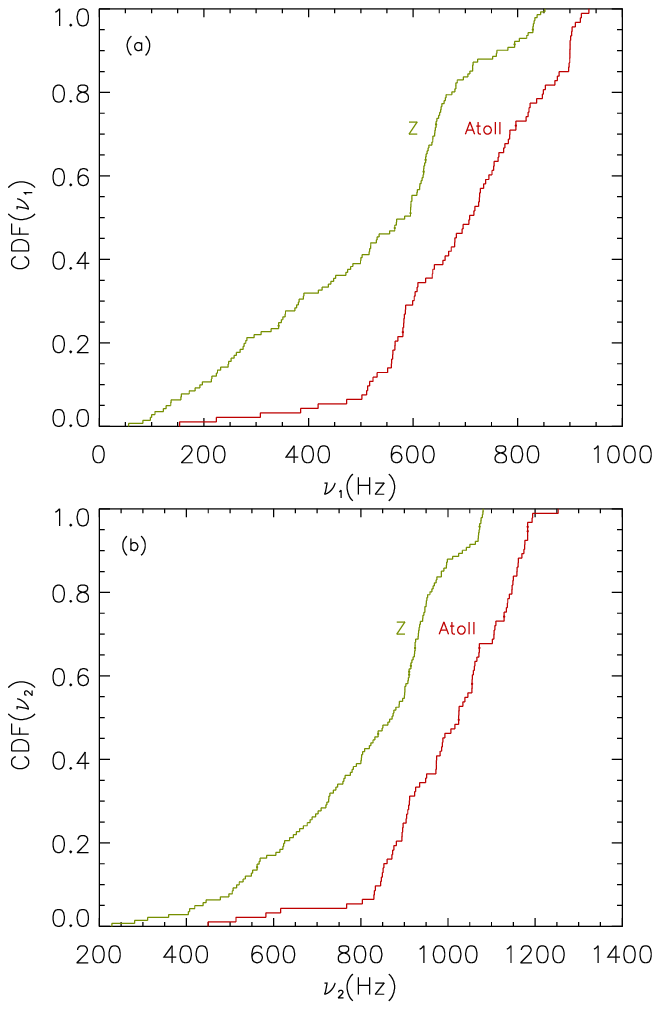}
\includegraphics[width=8cm]{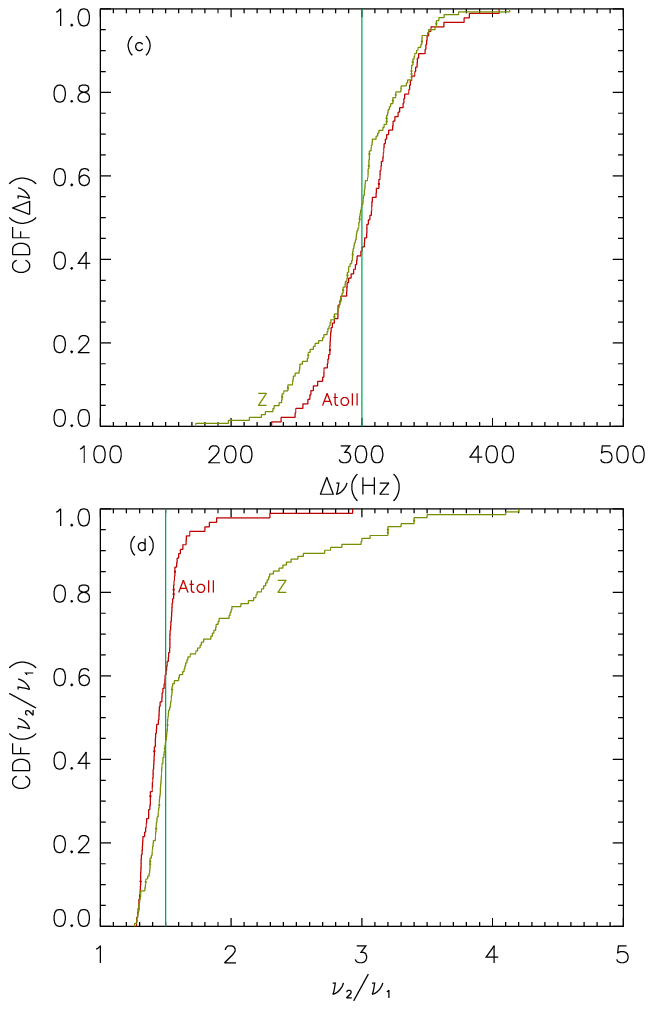}
\caption{(a)-(d) present the cumulative distribution function curves of $\nu_1$, $\nu_2$, $\Delta\nu$ and $\nu_2/\nu_1$. The data obtained without shift-and-add. The line of $\Delta\nu=300$ and $\nu_2/\nu_1=1.5$ are also plotted in (c) and (d), respectively.
}
\label{without shift-and-add}
\end{figure*}
\begin{figure*}
\includegraphics[width=16cm]{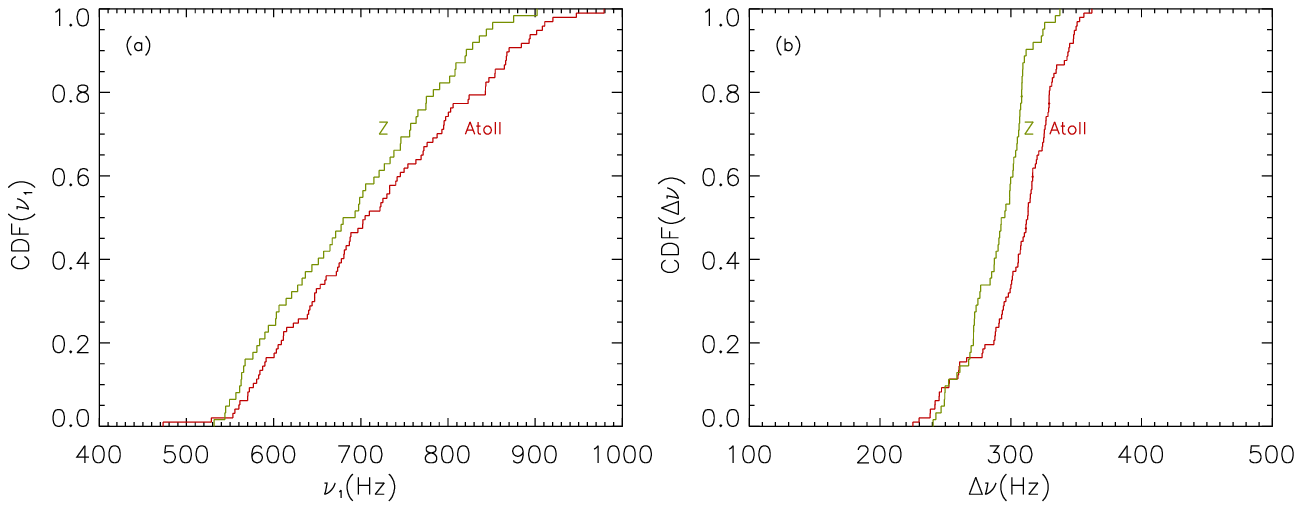}
\includegraphics[width=16cm]{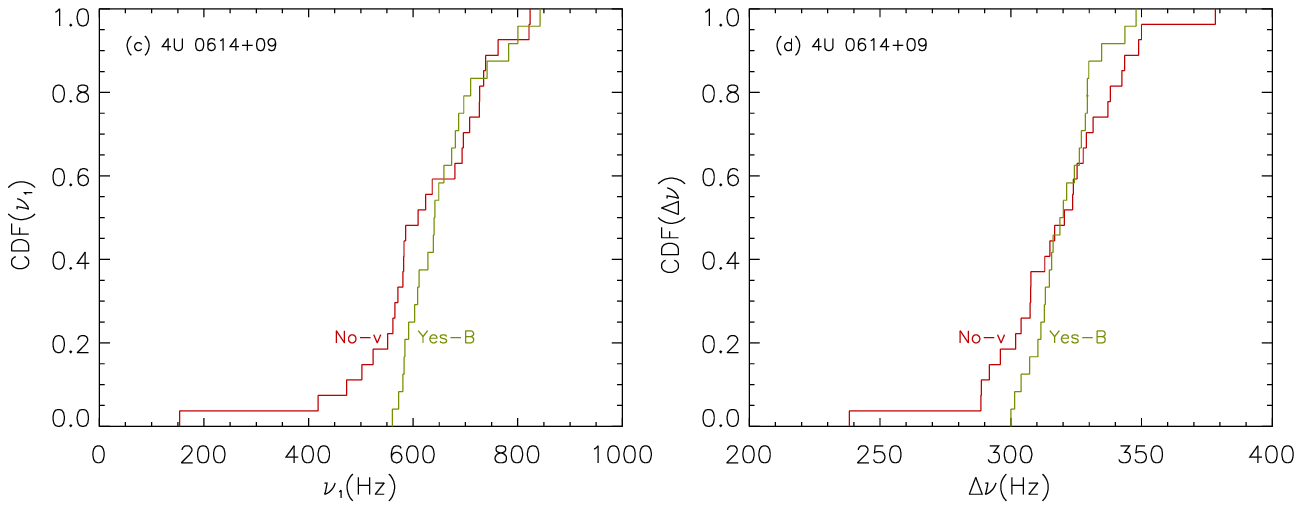}
\includegraphics[width=16cm]{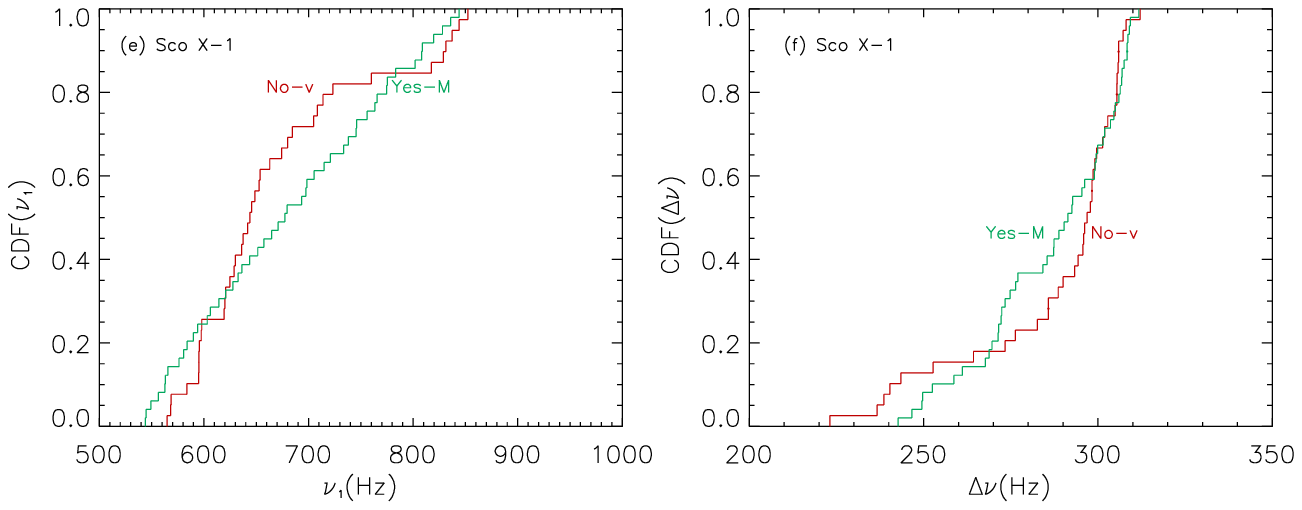}
\caption{(a)-(b) present the cumulative distribution function curves of $\nu_1$ and $\Delta\nu$, in which the data obtained with shift-and-add. (c)-(d) present the results of 4U 0614+09, where Yes-B means data with shift-and-add and from \citealt{Boutelier09}, No-v means data without shift-and-add and from \citealt{van Straaten00}. (e)-(f) are similar to (c)-(d), but data from \citealt{Mendez00} and \citealt{van der Klis97}, respectively.}
\label{with shift-and-add}
\end{figure*}

According to whether adopting shift-and-add technique, we separate the samples into two groups, the detail information of these sources
are shown in Table 1 and 2. Table 1 presents the results without shift-and-add technique,
which includes 93 data of Atoll sources and 141 of Z sources. Some of the results are obtained from figures, which are only kept with the integer parts. In case the data has the different up and down errors, we take the bigger one of them as its error, and if the data does not have error, we refer to the mean error of the same source and same observational method as its error. Results from the different authors are usually adopted by the different confidence ranges ($\Delta\chi^2=1$, $\Delta\chi^2=2.7$ or both, some authors even do not give the confidence range). For a conservative sense, we take $\Delta\chi^2=1$ (1 $\sigma$ single parameter) for all the data to calculate the weighted mean value (with 1 $\sigma$ confidence) of $\nu_1$, and $\nu_2$ (e.g. $\langle\nu_1\rangle$ and $\langle\nu_2\rangle$), then we reserve the digit according to the data of that source. Table 2 is similar to Table 1, but without shift-and-add technique, which does not reconstruct the distribution of pairs of frequencies \citep{Abramowicz03a, Belloni05}, and only the frequency difference is
meaningful \citep{Jonker00a}. So, we only show $\nu_1$, $\Delta\nu$ and their weighted mean values (with 1 $\sigma$ confidence, $\langle\nu_1\rangle$, $\langle\Delta\nu\rangle$), that includes 97 data of Atoll sources and 62 of Z sources. For 4U 1702-43 and 4U 1728-34, there are no errors of $\nu_1$ for all the data, so we only give their mean values.
It can be seen from Table 1 and 2 that five sources have the data both with and without shift-and-add. To be clear, we show the ranges of $\nu_2$ and $\Delta\nu$ as well as their mean values of these sources in Table 3.

It is noted that XTE J1701-462 and Cir X-1 are the two special
sources. XTE J1701-462 shows the link between "Z-track",
"$\nu$-track" and Atoll behavior in its CCD and HID diagrams
\citep{Lin09,Homan10}, and the twin kHz QPOs were detected when
this source was in Z phase \citep{Sanna10}. Cir X-1 was detected
transition between Atoll and Z source \citep{Soleri09} and this
source was also detected twin kHz QPOs when it showed property
much like Z source \citep{Boutloukos06,Boutloukos08b}.

\subsection{Analysis for the centroid frequency}

For the data without shift-and-add, the statistical results of kHz QPO frequency are as follows:
For Atoll (Z) sources, the weighted mean values (with 1 $\sigma$ confidence) of $\nu_1$, $\nu_2$, $\Delta\nu$ and $\nu_2/\nu_1$ are $744\pm10$\,Hz, $937\pm12$\,Hz\footnote{We get $571\pm16$\,Hz when calculating the weighted mean value of $\nu_2$, in which only two data point below this value. Considering that the weights of this two data may be over estimated, we replace them with mean weight of Atoll sources and the recalculated result is $937\pm12$\,Hz.},
$303\pm3$\,Hz and $1.42\pm0.01$ ($537\pm16$\,Hz, $886\pm11$\,Hz, $302\pm2$\,Hz and $1.47\pm0.01$).
The maxima of $\nu_1$ and $\nu_2$ in Atoll sources are 936 Hz and 1253 Hz (SAX
J1750.8-2900) while the corresponding maxima in Z sources are
852 Hz and 1081 Hz (Sco X-1), respectively.
For the data without shift-and-add, we take the mean errors of Atoll sources as the errors of 4U 1702-43 and 4U 1728-34, then
obtain the results (with 1 $\sigma$ confidence): for Atoll (Z) sources, the mean of $\nu_1$ and $\Delta\nu$ are $791\pm7$\,Hz and $311\pm3$\,Hz ($837\pm6$\,Hz and $293\pm2$\,Hz), respectively.
Table 3 shows that most of the maxima and mean values of $\nu_2$ of data with shift-and-add are bigger than those without shift-and-add. But we are not sure whether it results from the observational methods.
\begin{figure*}
\includegraphics[width=16cm]{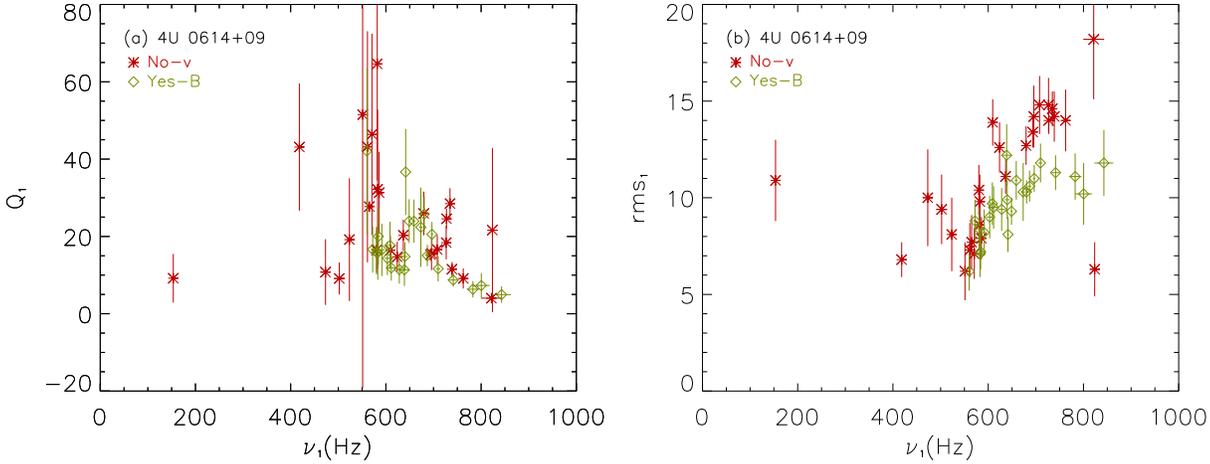} \\
\caption{
(a)-(b) present the plots of $Q_1~vs.~\nu_1$ and $rms_1~vs.~\nu_1$ for 4U 0614+09.
No-v means data without shift-and-add and from \citealt{van Straaten00}, Yes-B means data with shift-and-add and from \citealt{Boutelier09}.
}
\label{Q_rms}
\end{figure*}
We adopt K-S test to compare the centroid frequency distributions of kHz QPO between Atoll and Z sources as shown in Table 4, where the significance level $\alpha=0.05$ for all the tests (the two groups data have the different distributions if the test p-value is less than $\alpha$). The results obtained from data with and without shift-and-add are quite different:
for the data without shift-and-add, K-S test indicates that $\nu_1$, $\nu_2$ and $\nu_2/\nu_1$ of Atoll sources have the different distributions from those of Z sources, but $\Delta\nu$ show the consistent distribution between the two types of sources (see Table 4 and Fig.\ref{without shift-and-add}). On the contrary, K-S test from the data with shift-and-add show that the two types of sources have the consistent distribution of $\nu_1$, but inconsistent distribution of $\Delta\nu$  (see Table 4 and Fig.\ref{with shift-and-add} (a)-(b)). If we take the data of
the two observational methods together, K-S test shows that $\nu_1$, $\nu_2$, $\Delta\nu$ and $\nu_2/\nu_1$ of Atoll sources are all different from those of Z sources. It is obviously to see that the p-value of K-S test of $\Delta\nu$ changes sharply when adding the data with shift-and-add into the data without it (see Table 4). In order to find the reason that causes the different results, we select two sources: 4U 0614+09 (\citealt{van Straaten00}, 27 data; \citealt{Boutelier09}, 24 data) and Sco X-1 (\citealt{van der Klis97}, 39 data; \citealt{Mendez00}, 49 data), in which there exist the data with and without shift-and-add. The K-S test of these two sources show the consistent distributions of $\nu_1$, $\Delta\nu$ between the two different methods (see Table 4 and Fig.\ref{with shift-and-add} (c)-(f)). When we compare the distributions of data from the different methods with all the data, we find that the distributions of $\nu_1$, $\Delta\nu$ from the two methods are consistent in Atoll sources, but inconsistent in Z sources (see Table 4). There is only one Z source (Sco X-1)
with shift-and-add, which has the relatively bigger $\nu_1$ and smaller $\Delta\nu$ than other Z sources (see Table 2 and Fig.\ref{with shift-and-add} (a)-(b)). So, this selection bias, as well as the abundant data of Sco X-1, cannot reconstruct the frequency distribution of all Z sources, which may cause the different results. We also show the line $\Delta\nu=\rm constant$ (300\,Hz) and $\nu_2/\nu_1=1.5$ in Figure \ref{without shift-and-add} (c)-(d) respectively. K-S test indicates that $\Delta\nu$ is far from a constant distribution, so is $\nu_2/\nu_1$ (see also Table 4).

\subsection{Analysis for the quality factor and the rms amplitude}

As an extension of studying the influence of the two observational methods on kHz QPOs, We test the abrupt drop phenomena in $Q_1~vs.~\nu_1$ and $rms_1~vs.~\nu_1$ plots. We select the source 4U 0614+09 (\citealt{van Straaten00} and \citealt{Boutelier09}, see also Table 1 and 2) as a sample, where the data are obtained by the techniques with and without shift-and-add. We show results in Figure \ref{Q_rms}, where the drop in $Q_1~vs.~\nu_1$ and $rms_1~vs.~\nu_1$ plots can be seen if combining the two group data together, but the errors are too big to make a robust conclusion.

\section{Spin analysis in NS-LMXBs}

\begin{figure}
\centering
\includegraphics[width=8cm]{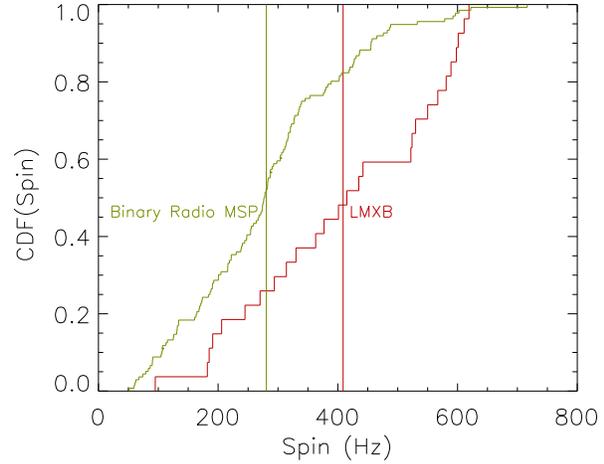}
\caption{NS spin frequency cumulative distribution function curves of 28 LMXBs and 136
radio binary MSPs. Their mean value (408\,Hz and  281\,Hz, respectively.) are also plotted.}
\label{spin}
\end{figure}

In the samples, 28 sources have the inferred NS spins from the
periodic or nearly periodic X-ray oscillations
\citep{Boutloukos08a}, where 9 sources are the accretion-powered
millisecond pulsars, 22 sources are
nuclear-powered millisecond pulsars, 3 sources are intermittent accretion-powered
oscillations pulsars \citep{Lamb09}, and 4 sources have been
detected both accretion-powered and nuclear-powered oscillations
and 2 sources have been detected both intermittent
accretion-powered and nuclear-powered oscillations.
The detail information of NS spins are listed in Table 5, where
one can see that all Z sources have not yet been detected the
inferred NS spins. The spin of XTE J1739-285 (1122\,Hz) has not yet been confirmed (see
\citealt{Kaaret07}), so we neglect this data when analyzing the result.
Figure \ref{spin} shows the NS spin frequency cumulative distribution function (CDF) curves of LMXBs, and the
range is from 95\,Hz to 619\,Hz with the mean value 408\,Hz.
Considering that NS in LMXB undergoes the spin-up process and will form a radio millisecond pulsar \citep{Bhattacharya91}, we
also analyze the NS spin frequency of 136 binary
radio millisecond ($>50$\,Hz or $<20$\,ms) pulsars\footnote{The data comes from Australia Telescope National Facility (ATNF)
pulsar catalog.}, with the spin range of 52--716\,Hz (1.4--19.4\,ms) and the mean value of 281\,Hz. The CDF of which is also shown in Figure \ref{spin},
where we notice that the NS spin frequencies of LMXBs and binary
radio MSPs share the similar range. The K-S test result of these two types of NS spins is shown in Table 4, indicating that they share the different distributions. The mean value of NS spin in NS-LMXBs is
bigger than that in binary radio MSPs.

\begin{figure}
\centering
\includegraphics[width=8cm]{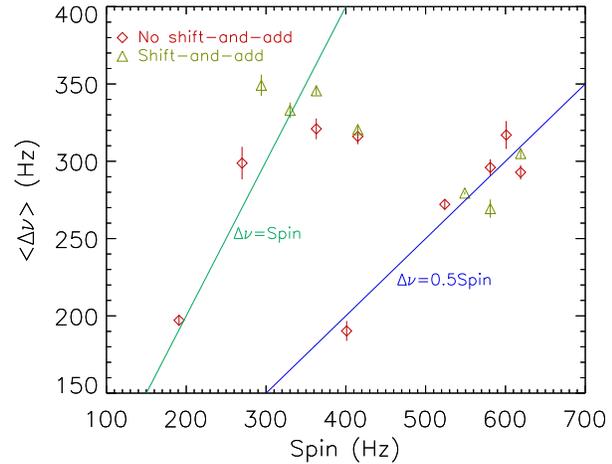}
\caption{$\langle\Delta\nu\rangle~vs.~\nu_s$ plot of 12 LMXBs. The solid
line represents $\langle\Delta\nu\rangle=\nu_s$. The diamond represents the data without shift-and-add, while triangle represents those with it.}
\label{spin_delt_nu}
\end{figure}

In the samples, 12 sources have both the detected twin kHz QPOs
and NS spins. We try to analyze the relation between peak
separations of twin kHz QPOs and their NS spin frequencies. Figure \ref{spin_delt_nu}
shows $\Delta\nu~vs.~\nu_s$ plot (both for data with and without shift-and-add). It can be seen that
there are two approximate clusters in the figure, one relates to the line of $\Delta\nu=\nu_s$ while the other relates to $\Delta\nu=0.5\nu_s$. These correlations are not obvious because of the large spread of the data.

\begin{table*}
\begin{minipage}{70mm}
\caption{NS spins in LMXBs}
\begin{tabular}{@{}lccl@{}}
\noalign{\smallskip}\hline\noalign{\smallskip}
Source (28) & $\nu_s$ & Type & Ref \\
 & (Hz) & & \\
\noalign{\smallskip}\hline\noalign{\smallskip}
Millisecond pulsars (9) \\
HETE J1900.1-2455 & 377 & I & 1;8 \\
IGR J00291+5934 & 598 & A & 1 \\
NGC 6440 X-2 & 206 & AN & 2 \\
SAX J1808.4-3658 & 401 & AN & 1 \\
XTE J0929-314 &185 & A & 1 \\
XTE J1739-285 &1122$^{\dag}$ & N & 3 \\
XTE J1751-305 & 435 & A &  1 \\
XTE J1807.4-294 & 191 & A & 1 \\
XTE J1814-338 & 314 & AN & 1 \\
\noalign{\smallskip}\hline\noalign{\smallskip}
Atoll Sources (13) \\
4U 0614+09 & 415 & N & 4 \\
4U 1608-52 & 619 & N & 1 \\
4U 1636-53 & 581 & N & 1 \\
4U 1702-43 & 330 & N & 1 \\
4U 1728-34 & 363 & N & 1 \\
4U 1915-05 & 270 & N & 1 \\
A 1744-361 & 530 & N & 1 \\
Aql X-1 (1908+005) & 550 & IN & 1;9 \\
IGR J17191-2821 & 294 & N & 1 \\
IGR J17511-3057 & 245 & AN & 5 \\
KS 1731-260 & 524 & N & 1 \\
SAX J1750.8-2900 & 601 & N & 1 \\
XB 1254-690 & 95 & N & 6 \\
\noalign{\smallskip}\hline\noalign{\smallskip}
Other Sources (6) \\
EXO 0748-676 & 522 & N & 7 \\
GS 1826-238 & 611 & N & 1 \\
MXB 1659-298 & 567 & N & 1 \\
MXB 1743-29 & 589 & N & 1 \\
SAX J1748.9-2021 & 442 & IN & 1;10 \\
SWIFT J1756.9-2508 & 182 & A & 1 \\
\noalign{\smallskip}\hline\noalign{\smallskip}
\end{tabular}
\end{minipage}
\begin{tabular}{@{}l@{}}
\begin{minipage}{80mm}
$\dag$: NS spin frequency has not yet been confirmed; \\
A: accretion-powered millisecond pulsar;\\
N: nuclear-powered millisecond pulsar;\\
I: intermittent accretion-powered oscillations pulsar;\\
1. Reference in \citealt{Boutloukos08a}; \\
2. \citealt{Altamirano10c}; \\
3. \citealt{Kaaret07};\\
4. \citealt{Strohmayer08a};\\
5. \citealt{Altamirano10b};\\
6. \citealt{Bhattacharyya07};\\
7. \citealt{Galloway10};\\
8. \citealt{Galloway07};\\
9. \citealt{Casella08};\\
10. \citealt{Gavriil07}, \citealt{Altamirano08a}, \citealt{Patruno08}.\\
\end{minipage}
\end{tabular}
\end{table*}
\section{Discussions and Conclusions}

From the kHz QPO data published before 2012, the following statistical results are obtained below:

(1). The K-S test results from data without shift-and-add technique show the inconsistency of $\nu_1$, $\nu_2$, $\nu_2/\nu_1$ distributions between Atoll and Z sources, which may result from the different properties of the two types of sources. The result also show the consistency of $\Delta\nu$ distributions, which indicates the twin kHz QPO of Atoll and Z sources to be the same physical origins. It can be seen from Figure \ref{without shift-and-add} and Table 3 that the distribution of $\Delta\nu$ is quite different from the prediction of beat model of twin kHz QPOs, so is $\nu_2/\nu_1$ different from the constant ratio (3:2). The similar conclusion is
also noticed by \citet{Belloni05}.

(2). From the results of data without shift-and-add, the weighted mean values of $\nu_1$ and $\nu_2$ of Atoll sources of
low luminosity are a little higher than those of Z sources of
high luminosity.
The maximum of $\nu_2$ in Atoll (Z) sources is 1253\,Hz (1081\,Hz)
(see Table 1), which is  the same order as the Keplerian orbital
frequency near the NS surface (for the NS with radius 15\,km and
mass 1.4 $M_\odot$, see \citealt{Zhang04}), so the kHz QPOs of
Atoll sources  could occur, usually, closer to the NS surface than
those of Z sources.

(3). For 4U 0614+09 and Sco X-1, the K-S test result shows the $\nu_1$ and $\Delta\nu$ distributions of the different observational methods are consistent, which may imply that shift-and-add technique can reconstruct the distribution of $\nu_1$ and $\Delta\nu$. The K-S test results of $\nu_1$ and $\Delta\nu$ between Atoll and Z sources from data with shift-and-add are quite different from those without it, and we think that there is only one Z source (Sco X-1) with shift-and-add, which has the relatively bigger $\nu_1$ and smaller $\Delta\nu$ than those of the other Z sources.
So, this selection bias, as well as the abundant data of Sco X-1, may cause the different results.

(4). We test the abrupt drop phenomenon in
$Q_1 vs. \nu_1$ and $rms_1 vs. \nu_1$ plots with data from different observational methods. We find that 4U 0614+09 shows the drop in $Q_1 vs. \nu_1$ and $rms_1 vs. \nu_1$ plots if combining the data with and without shift-and-add. But the errors are too big to make a certain conclusion.

(5). Considering that NS in LMXB undergoes the spin-up process and will form a radio millisecond pulsar \citep{Bhattacharya91}, NS spins in two systems may have a correlation, so we compare the NS spin frequency distribution in LMXBs with that in the binary radio MSPs (see Fig.\ref{spin}).
The range of NS spin frequencies in LMXBs (95-619\,Hz) is similar
to that of binary  radio MSPs (52-716\,Hz), but the spin mean value
in NS-LMXBs (408\,Hz) is bigger than that in binary radio MSPs (281\,Hz).
We consider that this difference may be due to the lack of NS spin
data from Z sources, which infers that the mean NS spin frequency of Z sources is lower than 281\,Hz.
Therefore, in general,
it is implied that the NSs in Z sources may be younger than those
in Atoll sources.

(6). In addition, with 12 LMXBs that have both detected twin kHz QPOs and inferred spin frequencies, we analyze the relation
between $\Delta\nu$ and $\nu_s$. It can be seen that
there are two approximate clusters in Fig.\ref{spin_delt_nu}, which relate to the lines of $\Delta\nu=\nu_s$ and $\Delta\nu=0.5\nu_s$, respectively. The data show a large spread, but it suggests some possible correlation between $\Delta\nu$ and the NS spin frequency, which shows a approximate consistent with the beat frequency models \citep{Miller98,Lamb01,Lamb03}.

\acknowledgements We thank M. M\'endez, J. Wang, Z.B. Li and H.X. Yin for helpful discussions. This work is supported by National Basic Research Program of China (2012CB821800 and 2009\\CB824800),  National Natural Science Foundation of China NSFC(11173034, 11173024, 10773017, 10778716, 11203064, \\10903005, 11303047) and Fundamental Research Funds for the Central Universities.

%
%

%

\end{document}